\documentclass[aps,twocolumn,showpacs,nofootinbib]{revtex4-1}
\usepackage{graphicx}   
\usepackage{latexsym}   
\usepackage{enumerate}
\usepackage{color}
\usepackage[normalem]{ulem}
\usepackage{float}
\begin{document}

\title{Effects of the QCD phase transition on hadronic observables in relativistic hydrodynamic simulations of heavy-ion reactions in the FAIR/NICA energy regime}

\author{Christian Spieles and Marcus Bleicher}

\affiliation{Institut f\"ur Theoretische Physik, Goethe-Universit\"at, Max-von-Laue-Strasse
1, D-60438 Frankfurt am Main, Germany\\
Helmholtz Research Academy Hesse for FAIR, Campus Frankfurt, Frankfurt, Germany}

\begin{abstract}
We investigate hadronic particle spectra and flow characteristics of heavy-ion reactions in the FAIR/NICA
energy range of $1\ A{\rm GeV} \le E_{\rm lab} \le 10\ A{\rm GeV}$ within a relativistic
ideal hydrodynamic one-fluid approach. The particlization is realized by sampling the Cooper-Frye distribution for a grand canonical hadron gas on a hypersurface of constant energy density. Results of the hydrodynamic calculations for different underlying
equations of state are presented and compared with experimental data and
microscopic transport simulations. The sensitivity of the approach to physical model inputs
concerning the initial state and the particlization is studied. 
\end{abstract}

\maketitle

\section[]{Introduction}
The exploration of the phase structure of the theory of strong interaction or Quantum Chromodynamics (QCD) has been one of the major goals of relativistic nuclear physics during the last 30 years \cite{Bass:1998vz}. Especially the programs at the AGS (EOS collaboration \cite{Scharenberg:2001gx}) and at CERN's SPS (NA49 \cite{Grebieszkow:2009jr}, and Shine collaboration \cite{Turko:2018kvt}) and later the RHIC-BES program (STAR collaboration \cite{Adam:2020unf}) have tried to to find unambiguous  signals for the onset of deconfinement. On the experimental side this search will continue in the next years with the novel facilities in Darmstadt and Dubna, namely the FAIR project \cite{Senger:2020iki} and the NICA project \cite{Kekelidze:2017ual}.  One the theoretical side, the search for the onset of deconfinement is plagued by the lack of quantitative predictions and high quality numerical simulations for a collision with a QCD phase transition \cite{Arsene:2006vf}. While this might sound surprising, it is unfortunately a fact that most transport simulations for heavy ion collisions in the FAIR/NICA regime do not allow to include a phase transition and can therefore at best provide the background dynamics \cite{Ono:2019ndq} (a notable exception is \cite{Bratkovskaya:2011wp}). 

In contrast, relativistic hydrodynamics simulations can provide new insights by incorporating a phase transition at finite baryo-chemical potential as needed for this energy. The application of hydrodynamic models to the simulations of nuclear collisions has a long history \cite{Belenkij:1956cd,Amsden:1975yg,Wong:1975fyg,Csernai:1980sc,Mishustin:1987aq}. The strength of this approach lies in the fact that apart from the basic model assumption of local thermal equilibrium essentially only the choice of a concrete equation of state enters as a physical input. 

At low energies, the hydrodynamic picture of a single fluid describing the
interaction of projectile and target nuclei has been used early on to study collective effects like directed flow and the dependence
of these effects on the nuclear equation of state (see, e.~g.
\cite{Wong:1975fyg,Csernai:1980sc,Rischke1995c}). 
Spectra of secondary particles, however, have rarely been analyzed in a
purely hydrodynamic description of low energy heavy ion collisions, a notable
exception being the two-fluid model approach of \cite{Russkikh1994}.

At high collision energies, on the other hand, hydrodynamic models have been
found appropriate for the description of the hot and essentially baryon free matter created at top RHIC and LHC energies.
Especially since the start of the experimental program at RHIC in 2000, one-fluid hydrodynamics has seen a tremendous increase in its applications to relativistic heavy ion collisions \cite{Huovinen:2001cy,Zhou:2020pai}.

In addition to pure one-fluid hydrodynamics, alternative approaches have been developed, namely two- and three-fluid hydrodynamics and Boltzmann+Hydro hybrid models. Here, the initial stage consisting of the target at projectile nucleons is treated separately from the created fireball. This allows to include the strongly anisotropic momentum distribution of the initial stage and to include a phase transition for the central reactions stages. Especially at intermediate energies where the initial baryon currents can be separated in momentum space such approaches are very valuable. Prime examples for three fluid hydrodynamics models are \cite{Katscher:1993xs,Brachmann:1997bq,Ivanov:2005yw,Batyuk:2016qmb}, while the idea of coupled Boltzmann+Hydro simulations was pioneered in \cite{Dumitru:1999sf,Bass:1999tu,Teaney:2001av,Steinheimer:2007iy,Petersen:2008dd} and later applied in \cite{Werner:2010aa,Song:2011qa,Denicol:2018wdp}.

A major drawback of hybrid and multi-fluid approaches is that they typically relay on a separation (in space and momentum space) of the baryon currents. Especially at low energies this assumption might not be fulfilled and the different currents (propagated in different fluids) should be treated as a single thermodynamic system at each space time point. The typical minimal scale from which on a simple multi-fluid approach without unification of the fluids can be used is provided by the thermal velocity in a typical cell in relation to the Moeller velocity between the fluids (or baryon currents). This was studied in \cite{Brachmann:1997bq} and indicates that even up to 11 AGeV one-fluid hydrodynamics provides a good approximation to the expansion stage of the reaction. 

In the present study, a relativistic ideal one-fluid model of heavy-ion collisions is applied to the FAIR/NICA energy regime. The
reaction dynamics is described hydrodynamically starting from cold projectile and target nuclei up to the
particlization into hadrons. The final $m_t$ spectra of protons, pions and
kaons from central Au+Au collisions are systematically analyzed and compared with experimental data
for different equations of state and for different beam energies. 
Excitation functions of the flow parameters
$v_1$ and $v_2$ in mid-central collisions are also studied. A special focus
is laid on the importance of initial state density fluctuations for these
observables.

\section{The model}
\subsection{Initialization}
The model can either be initialized with averaged nuclear profiles of the nucleons in each projectile and target nucleus or using a Monte Carlo procedure to provide fluctuating initial stages for event-by-event simulations. 
Consequently, in the 'event-by-event' mode multiple hydrodynamic events with different initial states are
calculated in order to obtain a sample of final states which are then
averaged for the present study. While in the 'averaged' mode initial configurations are first averaged and the a single hydrodynamic evolution is performed. Thus, the 'averaged' mode neglects initial stage fluctuations and leads in general to smoother initial density and energy profiles. The practical implementation of the fluid initialization follows
\cite{Petersen2008}: The spatial coordinates of the nucleons from an
individual initial state is replaced by three-dimensional
Lorentz-contracted Gaussians of width $\sigma=1$~fm. The resulting sum of
the corresponding energy, momentum and baryon-densities of all nucleons is then mapped on
the spatial grid with cell-size $(0.2\ \rm fm)^3$.

\begin{figure}[t]
\includegraphics[width=0.55\textwidth]{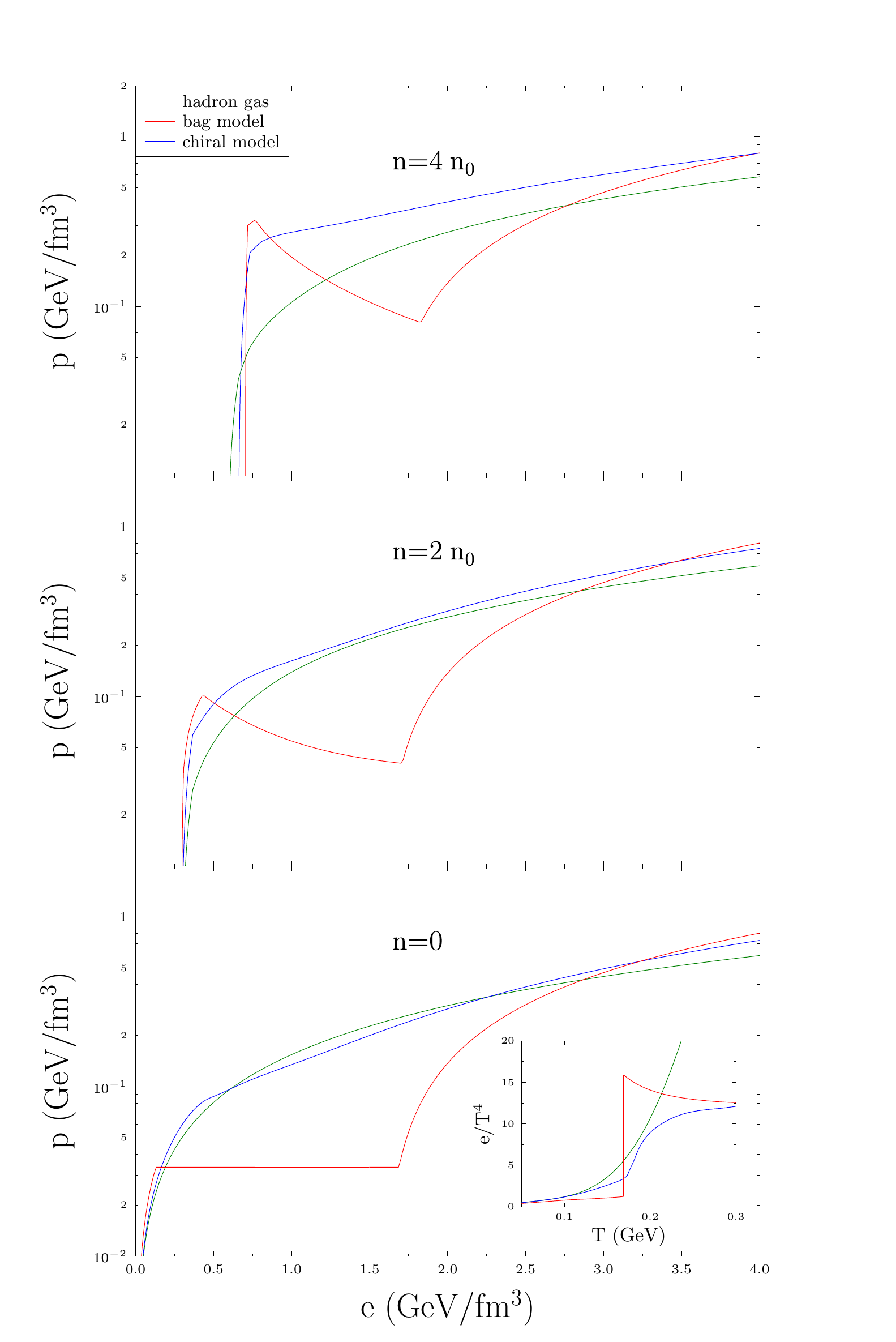}
\caption{
Equations of state employed in the hydrodynamic model simulations.  Shown is the pressure $p$
as a function of energy density $e$ for different fixed values of net-baryon density
$n$ ($n_0$ is the net-baryon density of the nuclear ground state). 
 The inset in the bottom figure displays the dimensionless quantity $e/T^4$
as a function of $T$ for the case $n=0$. 
\label{fig:plot_eos} }
\end{figure}

\subsection[]{Equations of motion}
Ideal relativistic hydrodynamic assumes conservation of energy, momentum and charges:
\begin{equation}
\partial_\mu T^{\mu\nu}=0 \quad {\rm and}\quad \partial_\mu N^{\mu}=0
\end{equation}
Here, $T^{\mu\nu}$ is the energy-momentum tensor and $N^\mu$ is the baryon four-current.
These quantities can be expressed in terms of the fluid's four-velocity $u^\mu$ and the
thermodynamic state in the local restframe of the fluid, described by the energy density $\epsilon$,
the pressure $p$ and the baryon density $n$:
\begin{equation}
T^{\mu\nu}=(\epsilon +p) u^\mu u^\nu - p g^{\mu\nu} \quad {\rm and} \quad 
N^{\mu}=n u^\mu
\end{equation}
In addition to the hydrodynamic equations, a specific equation of state of the matter
is required in the form $p=p(\epsilon,
n)$. The following coupled equations then determine the time evolution of the system: 
\begin{eqnarray}\label{eq:eom}
\partial_t T^{00} + \vec{\nabla} \cdot (T^{00} \vec{v}) &=& -\vec{\nabla}\cdot (p\vec{v})  \\
\partial_t T^{0i} + \partial_i (\sum{T^{0j} v_j }) &=& - \partial_i p \quad {\rm for}\quad  
i=1,2,3\\
\partial_t N^0 + \vec{\nabla} \cdot (N^0 \vec{v}) ) &=& 0
\end{eqnarray}

These equations are numerically solved with the
SHASTA algorithm \cite{Rischke1995a,Rischke1995b}. In each time step, the algorithm consistently links the 
discretized conserved quantities $T^{00}$, $T^{0i}$  and $N^0$ (given in the calculational
frame) with the pressure $p$ (given in the 
local rest frame)  and the fluid's three-velocity $\vec{v}$ of each cell.

A main advantage of the SHASTA as compared to other approaches is its ability to handle shock wave formation and propagation by the flux corrected transport with non-linear feedback \cite{handbook_of_shockwaves}. Especially for our application at low energies, this is a crucial feature, because the entropy production in the initial stage is completely described by shock wave formation.

\subsection[]{Equations of state}
The Equation of State (EoS) is the main ingredient into the hydrodynamic simulation. At high energies, i.e. for systems with (nearly) vanishing net-baryon density the EoS can be obtained from fits to lattice QCD calculations  \cite{Huovinen:2009yb,Monnai:2019hkn}, at low energies with high baryo-chemical potentials lattice QCD calculations can unfortunately not provide ab-initio results for the EoS and one has to rely on phenomenological approaches that converge to lattice QCD results in certain limits. For the present study, different equations of state have been assumed: The {\it hadron gas} is an 
ideal relativistic quantum-gas of known baryon and meson resonances up to $\approx 
2\ A\rm GeV$. It reflects the same degrees of freedom as the hadronic cascade model UrQMD. For 
details, we refer to \cite{Zschiesche2002}.
The {\it MIT bag model} equation of state and its implication on fluid-dynamic properties have been 
discussed in \cite{Rischke1995b}. This (rather schematic) equation of state assumes a version of the $\sigma-\omega$-model for the 
hadronic phase and a non-interacting gas of massless $u$ and $d$ quarks and gluons confined by a 
bag pressure for the QGP phase. The transition between the two phases is of first-order, governed 
by the Gibbs condition. The bag model EoS serves mainly to show case the most drastic effects of a phase transition. According to the {\it chiral model} \cite{Steinheimer2011}, strange and non-strange 
baryons interact via mesonic mean-fields. This model equation of state exhibits a cross-over phase transition to 
deconfined matter. 

Fig.~\ref{fig:plot_eos} depicts the three equations of
state for three different fixed values of net-baryon density $n$. Shown is
the pressure as a function of energy density. This is the hydrodynamically
relevant representation of the equation of state, since the pressure can be
regarded as the 'response' of the matter for a given condition (baryon
density and energy density) the system is forced into in the course of the
evolution. Obviously,
the different model equations of state indicate strongly differing
thermodynamic behaviour of nuclear matter. It
will be shown below, how this reflects in the outcome of the different hydrodynamic simulations.

\subsection[]{Hadronic freeze-out}
The transition between the fluid phase and the microscopic transport description of the reaction is 
given by the Cooper-Frye equation \cite{CooperFrye1974}:
\begin{equation}
E\frac{dN_i}{d^3p}=\int_\sigma f_i(x,p) p^\mu d\sigma_\mu \quad .
\end{equation}
$f_i$ denotes the particle distribution function of hadron species $i$,
boosted according to the fluid velocity of the
hypersurface element.

The hydrodynamic freeze-out or particlization takes place on a hypersurface of equal local energy density
$\epsilon_{\rm fr}$ of the 
fluid. The numerical procedure for determining the hypersurface is described in \cite{Huovinen2012}.
In the present study, reaching nuclear ground state energy density is used as the
particlization criterion, i.~e. the value of local energy density where the hydrodynamic fields are converted into
hadrons is $\epsilon_{\rm fr}=\epsilon_0$. 

Hadrons are sampled according to the grand-canonical description
of a non-interacting relativistic hadron gas in equilibrium, taking into account the same
degrees of freedom as the microscopic transport model UrQMD. The values of
$T$ and $\mu_B$ depend on the baryon density of the fluid at the particular hypersurface
element. Different to the implementation of \cite{Petersen2008}, the
 conservation of
 energy and baryon number in the grand-canonical ensemble  refers to
expectation values, not to individual events. Electric charges are not propagated
explicitly in the hydrodynamic simulation. In order to recover the net-charge
of the initial state of the heavy-ion collision in the final state, the
$Z/A$ ratio of the system is used as  global constraint which implies a 
certain isospin asymmetry for each element of the freeze-out hypersurface,
depending on the local values of $T$ and $\mu_B$. The sampling
of particles then takes into account this isospin asymmetry.

In principle, there should be no need for an additional normalization of the
particle yields, since
the sampling procedure runs over all elements of the particlization
hypersurface. Indeed, the 'raw' model-outcome is consistent with baryon
number conservation on a $1\ \%$ level in the case of the hadron gas equation
of state and assuming an averaged initial state. Taking into account initial state fluctuations
reduces the agreement to a $5\ \%$ level. Equations of state other than the
ideal hadron gas in some cases lead to even larger apparent violations of baryon
number conservation. Therefore, we calculate an overall correction factor
for the net-baryon number expectation value of each individual particlization hypersurface and apply it in the sampling procedure (for all hadron species).

The transition from the hydrodynamic evolution is followed by final
state interactions (scattering and decays). For these, the UrQMD-cascade description is employed, as in
\cite{Petersen2008,Werner:2010aa,Song:2011qa,Denicol:2018wdp}. However, since the particlization is assumed to take place at significantly lower energy densities, the hadronic scatterings at
this late stage are not relevant for any observables analyzed in this study. 

\section{Results}
\subsection[]{Hadronic particle spectra}

\begin{figure*}[t]
\includegraphics[width=1.1\textwidth]{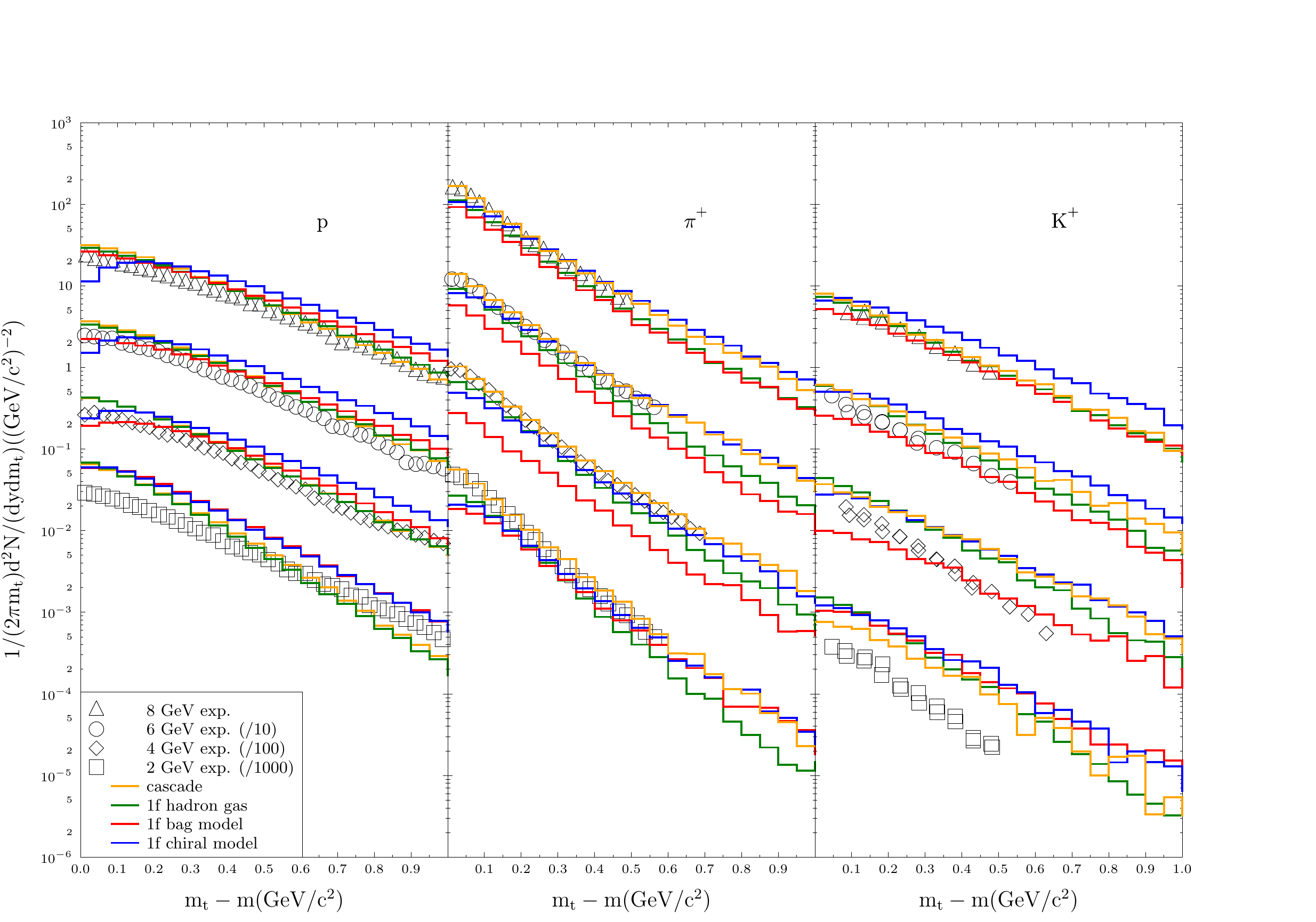}
\caption{Invariant particle yield at $|y-y_{cm}|<0.05$ as a function of $m_t-m_0$ 
in central Au+Au collisions at (2, 4, 6, and 8) $A$GeV. Shown are
experimental data for protons from E895 \cite{Klay2002}, for $\pi^+$ from
E895 \cite{Klay2003} and for $K^+$ from E866 and E917 \cite{Ahle2000} in comparison
with different hydrodynamic simulations and the UrQMD cascade model.
\label{fig:plot_mt} }
\end{figure*}

\begin{figure}[h]
\includegraphics[width=0.5\textwidth]{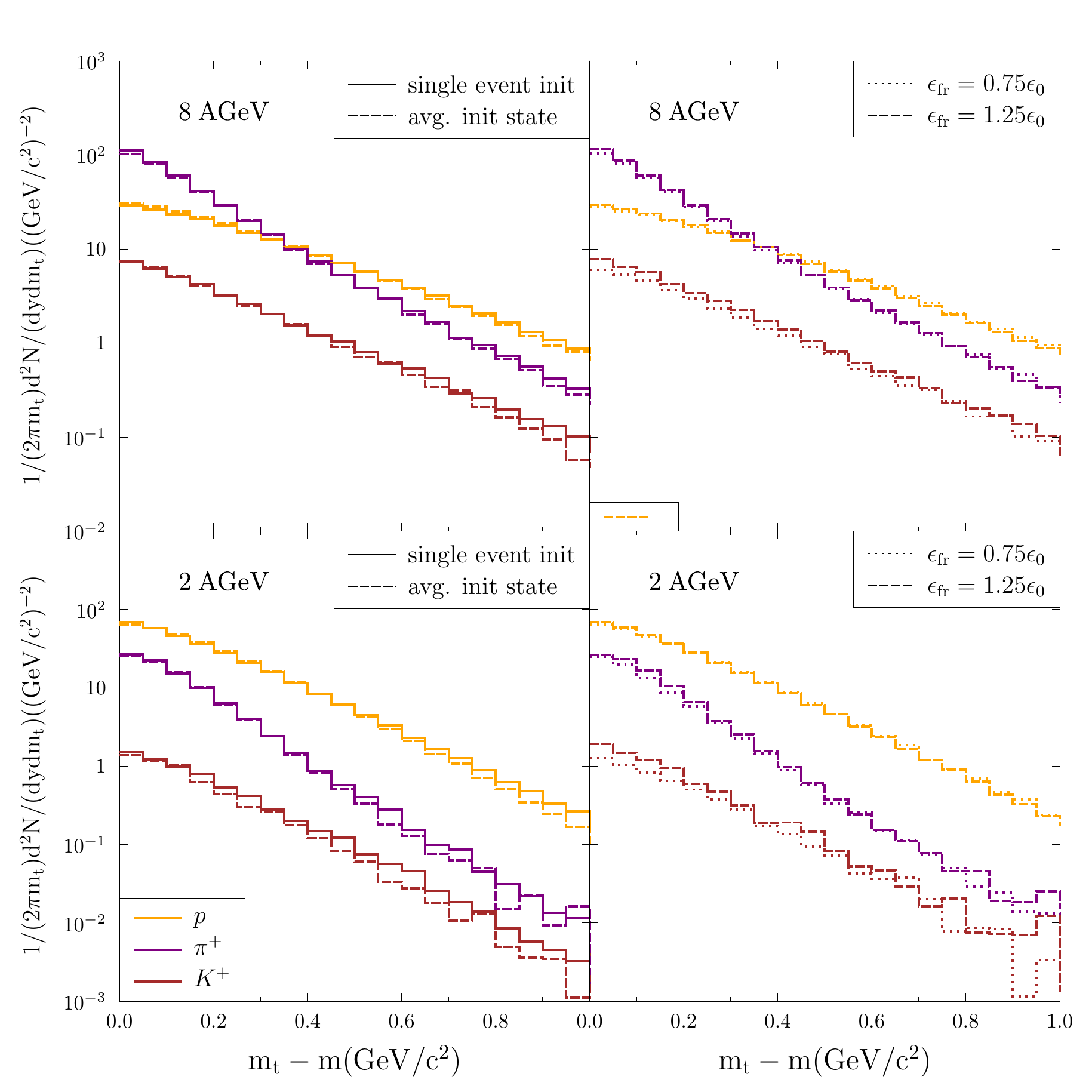}
\caption{Invariant particle yield at $|y-y_{cm}|<0.05$ as a function of $m_t-m_0$ 
in central Au+Au collisions at 2 and 8 $A$GeV according to the hydrodynamic
calculation with the hadron gas equation of state. Shown are the sensitivities
with respect to initial state density fluctuations (left hand side) and with
respect to the value of the particlization parameter $\epsilon_{\rm fr}$ (right hand side).
\label{fig:plot_sensi} }
\end{figure}

Figure~\ref{fig:plot_mt} shows transverse mass spectra of protons,
$\pi^+$ and $K^+$ at midrapidity for central Au+Au collisions at
different energies. In the simulations, the impact parameter is set
to $b=2$~fm, whereas the experimental data represent the $5$~\% most central
events. Hydrodynamic simulations using the three different equations of
state are compared with the microscopic UrQMD model and experimental data
from \cite{Klay2002}\cite{Klay2003}\cite{Ahle2000}. Firstly, one notes a close
similarity between the hydrodynamic simulation using the hadron gas equation of
state and the microscopic transport model for all hadron species at all
energies. This is a surprising fact, given that the hydrodynamic model comes with
extremely few free parameters, none of which is fitted to experimental data
from heavy ion reactions. The final state of the hydrodynamic description only
reflects local energy and momentum conservation in combination with the assumption 
of immediate local equilibration of all hadronic degrees of freedom at all
times. The microscopic model, on the other hand, goes to great length to
take into account the non-equilibrium physics of a nuclear collision by
implementing hundreds of elementary cross sections and decay
channels. However, the fundamentally different models show quantitatively very similar
amounts of nuclear flow (indicated by the slope of the protons), as well as entropy and
strangeness production (yield and slope of the pions and kaons,
respectively). 

There are some significant differences between the
simulations with different equations of state, though. Firstly, the
chiral model equation of state shows the strongest flow, indicated by
relatively flat slopes of protons, and also secondaries,  in particular at
higher energies. This corresponds to the relatively high values of
pressure at given baryon density in a wide range of energy densities, see
Fig.~\ref{fig:plot_eos}. The bag model exhibits interesting deviations from
the other scenarios, concerning the pion and kaon yields at
4~$A$GeV, which appear clearly suppressed. This can be understood in view of
the low pressure (and relatively low temperature) in the mixed phase, which
seems to dominate the reaction
dynamics just at 4~$A$GeV (at this energy also fluctuations due to spinodal instabilities reach their maximum \cite{Steinheimer:2012gc}). At 2~$A$GeV it is plausible that only a small
portion of the reaction volume enters the mixed phase of the equation of
state. 

The comparison with experiment shows that UrQMD and the hydrodynamic calculations with all inspected equations of state
do not fully capture the flow in the slope of the proton spectrum at 2~$A$GeV.  
At 4-8~$A$GeV, both the hydrodynamic models as well as the UrQMD model agree
nicely with the proton data, perhaps with the exception of the chiral model equation of state
which renders slightly too much transverse collective motion.
The spectra of pions are well reproduced by hydrodynamics at all energies. As mentioned above, the bag
model equation exhibits a remarkable deviation from the other model
scenarios at intermediate energies, most pronounced at 4~$A$GeV. Hence, the outcome
of the bag model model equation of state is also clearly in conflict with experimental data.

All models presented here fail to reproduce the kaon yield at 2~$A$GeV and
4~$A$GeV. 
It appears that the steep
rise of the $K/\pi$ ratio when comparing 4~$A$GeV collisions with 2~$A$GeV
poses a challenge for dynamical model descriptions of hadron production in
heavy-ion collisions.
Statistical models with an assumed energy dependence of the thermodynamical
parameters describe the observed strangeness production \cite{PBM2002}.
However, new complications arise for most models when trying to explain
experimental data at higher collision energies \cite{Alt2008}.

As for the present analysis, in particular the hadron gas equation of state shows very good
agreement with the experimental data for all hadron species at 6 and
8~$A$GeV. It seems that these collision energies with
significantly higher collision rates and hadron production cross
sections way above threshold are consistent with the assumption of thermal
and hadrochemical equilibrium during the determining reaction stage.

As mentioned above, apart from the equation of state, the hydrodynamic model
does not leave much space for different physical model parameters or
assumptions that could affect the outcome. In the following, the sensitivity of the
hadron spectra to this model input is analyzed. Firstly, one might
question the assumed initial state of the fluid as being determined by the distribution of a finite number
of nucleons on the computational grid. It could be argued that averaging
density fluctuations over an ensemble of projectile and target nuclei
constitute the appropriate initial state for the fluid dynamical
description. In Fig.~\ref{fig:plot_sensi}, the plots on the left hand side show an analysis
of the hadron spectra for the standard scenario with initial state density
flucutations in comparison with the scenario with an averaged initial state.
The resulting $m_T$ distribution of all particle species are only very weakly
sensitive to the choice of the initial state scenario ('averaged' initial conditions vs. 'event-by-event' initial conditions). This holds for the whole energy range explore in this study. 

Next we explore the transition criterion from the hydrodynamic stage to the Boltzmann dynamics. The parameter $\epsilon_{\rm fr}$, the value of the energy density that serves as the
hydrodynamic freeze-out or particlization criterion. For the present study, the value is set
to $\epsilon_{\rm fr}=\epsilon_0$, the energy density of the nuclear ground
state. This value is motivated by the typical mean free path of the nucleons and may be seen as a lower limit for this value. In contrast, in \cite{Petersen2008}, a value of  $\epsilon_{\rm fr}=5 \epsilon_0$ was
assumed for collisions in the highly relativistic regime. However, this is not appropriate for the present study where such an energy density may not even be reached as a peak value. 
To be consistent, the energy density where particlization takes place is
instead chosen such that cold nuclei are below or just at the threshold of
the fluid description. Much lower values of $\epsilon_{\rm fr}$, on the other hand, can
hardly be justified due to the resulting large mean free path or low
scattering rates. Thus, the reasonable parameter range  of $\epsilon_{\rm fr}$
is rather small for low energy nuclear collisions. Still, the sensitivity of the the model on its
variation should be checked. In Fig.~\ref{fig:plot_sensi}, the plots on the right hand side show an analysis
of the hadron spectra for $\epsilon_{\rm fr}=0.75 \epsilon_0$
in comparison with $\epsilon_{\rm fr}=1.25 \epsilon_0$. The spectra from the
hydrodynamic model calculations are very robust with respect to the changes of $\epsilon_{\rm
fr}$ in this range. One just observes a slight suppression of kaon
production for the lower value of $\epsilon_{\rm fr}$, which is
to be expected considering the lower average freeze-out temperatures associated with this choice of $\epsilon_{\rm fr}$. 

\begin{figure}[h]
\includegraphics[width=0.6\textwidth]{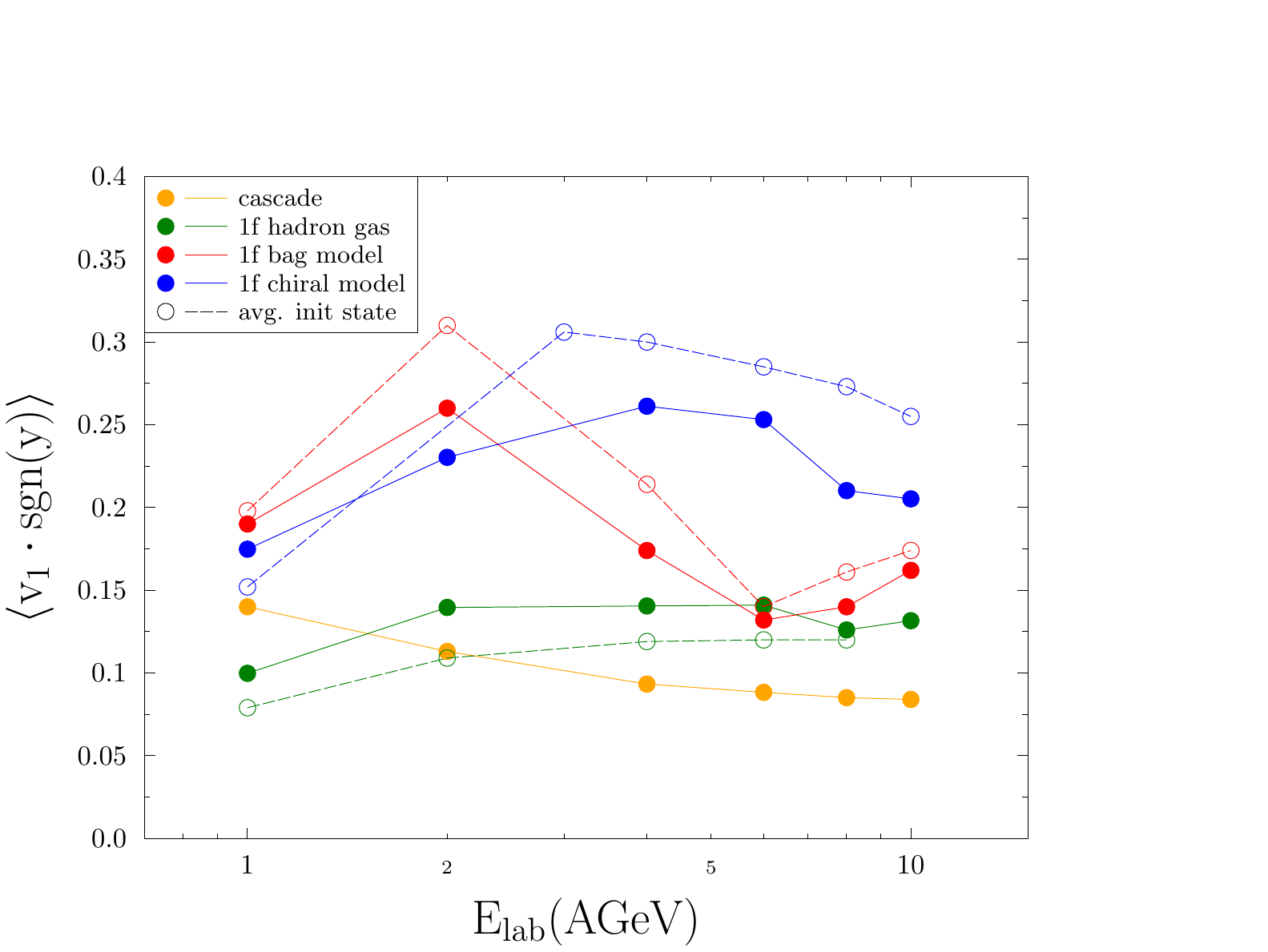}
\caption{Mean directed transverse momentum  in the reaction plane for protons
in  Au+Au collisions ($b=7$\ fm) at different energies. Shown are the
results of hydrodynamic  calculations for different equations of state with and without taking into account initial state fluctuations of the nuclear matter
distributions. Also shown is the result of the microscopic UrQMD transport
simulation.
\label{fig:plot_pxdir} }
\end{figure}

\subsection[]{Directed flow}
Excitation functions of the directed flow have been studied in \cite{Rischke1995c}
as a possible way of probing the equation of state of nuclear matter in
a fluid-dynamic model. These calculations did not take into account hadronic
freeze-out - as most of the early studies - and relayed directly on the analysis of the energy momentum tensor of the baryonic fluid. The proposed quantity for measuring the effect was the mean
$x$ component of the fluid momentum (with $x$ pointing transversely in the reaction-plane), integrated over
forward rapidity and weighted with the baryon number rapidity density. Unfortunately, this
is not an experimentally accessible observable. 

In the present study, the directed flow is calculated on the basis of the nucleons
after particlization and subsequent hadronic freeze-out from the final
microscopic transport stage. The directed flow is quantified by the first Fourier
coefficient of the azimuthal distribution of the nucleons with respect to
the event plane:
\begin{equation}
v_1=\langle \cos{(\phi))}\rangle= \left\langle \frac{p_x}{p_T}\right\rangle \quad .
\end{equation}
The value of  $v_1$ of particle distributions changes its sign at  midrapidity for symmetry reasons.
Therefore, in order to define a meaningful observable integrated over rapidity\footnote{An alternative observable may be defined by the slope of $v_1$ at midrapidity. However, the extraction of a slope parameter is less reliable than an integrated value.}%
, the quantitity $\langle v_1 {\rm sgn}(y)\rangle$, averaging over all individual
nucleons, is considered. This quantity, called the directed flow, is depicted in
Fig.~\ref{fig:plot_pxdir} as a function of beam energy for different model
equations of state. 

The UrQMD cascade calculation may serve as a baseline to which the different
hydrodynamic simulations are compared. The energy dependence of the
directed flow is rather weak. One observes a slow decrease which seems to
saturate at a value  lower than all fluid-dynamic model scenarios. For the hydrodynamical calculation with a 
hadronic equation of state with the same degrees of freedom as the UrQMD
model, the directed flow effect is also rather weak and the energy
dependence is also not very pronounced. While, this scenario exhibits a slow
increase of the directed flow parameter instead of a decrease, the numerical values of the directed flow are rather similar for both hadronic scenarios.

In contrast, the chiral model equation of state
shows a completely different energy dependence of the
directed flow: Firstly, the flow effect is stronger at all energies.
Secondly, it exhibits a pronounced maximum at around 4 $A$GeV. In contrast,
the bag model equation of state renders the minimum of the directed flow at
around 6 $A$GeV  
which perfectly coincides with the results of \cite{Rischke1995c}, where
this minimum had been proposed as a qualitative signal for the transition of
hadronic matter to quark and gluon degrees of freedom. The reason is, of
course, the extreme softening of the equation of state in the mixed phase,
as can bee seen from Fig.~\ref{fig:plot_eos}. It is notable that
the chiral model also incorporates a phase transition to deconfined matter
with quark degrees of freedom, which is best recognizable in the inset of  Fig.~\ref{fig:plot_eos}
as $e/T^4(T)$. However, assuming this equation of state does not lead to the
minimum in the directed flow in the inspected energy range.

As mentioned above, the values of the pressure for given baryon and energy
density determine the hydrodynamic evolution. In this respect, the
differences between the chiral model equation
of state and the bag model of state are much more pronounced than the
differences between the chiral model and the hadron gas over a considerable
range of energy density and baryon density. This explains the
differences of the directed flow parameter between 2 and 8 $A$GeV. For
low beam energies between 1 and 2 $A$GeV, on the other hand, the chiral
model and the bag model equation of state show very similar amounts of
directed flow, whereas the hadron gas differs considerably. This seems
plausible in view of the Fig.~\ref{fig:plot_eos} (centre and top), where
the pressure of nuclear matter for energy densities and baryon densities expected in
heavy-ion collisions at these energies can be read off. The hadron gas with
its numerous noninteracting degrees of freedom shows much lower values of
pressure than both the bag model and chiral model.

For the hydrodynamic model Fig.~\ref{fig:plot_pxdir} also shows
the resulting directed flow parameters according to the scenario with the 'averaged' initial
state, where density fluctuations of single events are averaged over. In this case, the effect of
directed flow is consistently stronger than in the standard scenarios. The
momentum correlations corresponding to the directed flow appear to be
distorted by random fluctuations of the spatial fluid distribution. The importance of the initial state fluctuations was also observed in \cite{Petersen:2010md}.

\begin{figure}[h]
\includegraphics[width=0.6\textwidth]{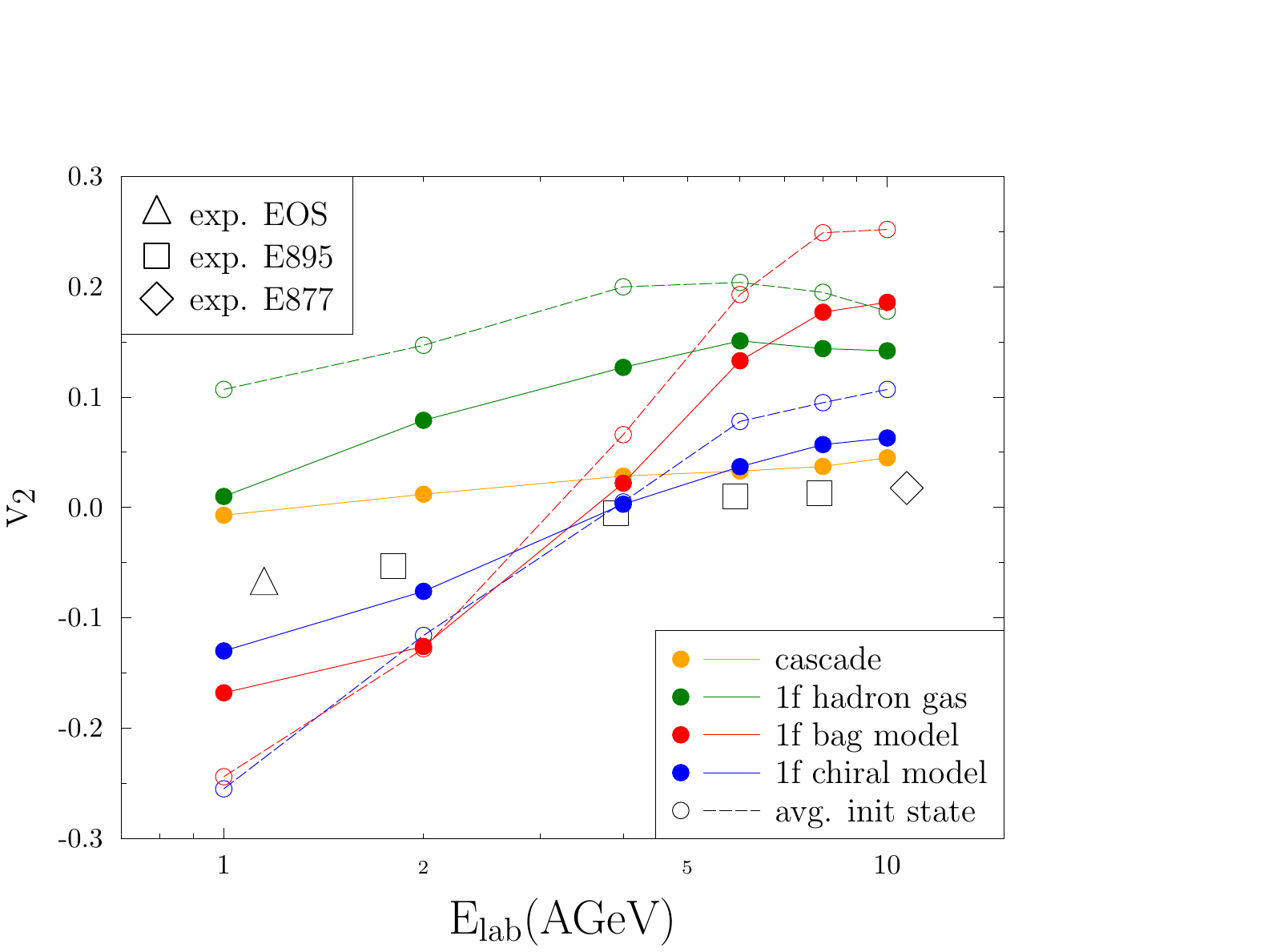}
\caption{Excitation function of the elliptic flow $v_2$ of protons at $|y-y_{cm}|<0.05$ 
in  Au+Au collisions ($b=7$\ fm). Shown are hydrodynamic calculations with
different equations of state, the results of the microscopic transport model UrQMD
and experimental data from \cite{Pinkenburg1999}. Also shown are the
hydrodynamic results without taking into account initital state density
fluctuations.
\label{fig:plot_v2_1} }
\end{figure}

\subsection[]{Elliptic flow}
Finally, we turn to the elliptic flow or $v_2$. The elliptic flow is quantified by the second Fourier
coefficient of the azimuthal distribution of the nucleons with respect to
the event plane:
\begin{equation}
v_2=\langle \cos{(2\phi))}\rangle= \left\langle \frac{p_x^2-p_y^2}{p_x^2+p_y^2}\right\rangle
\quad .
\end{equation}
The experimentally observed elliptic flow of protons at
midrapidity for mid-central Au+Au collisions in the beam
energy range between 1 and 10 $A$GeV exhibits  a characteristic sign change
\cite{Pinkenburg1999}. The elliptic flow is especially important, as it provides an extremely sensitive link to the Equation of State as discussed in \cite{Danielewicz:2002pu}. In Fig.~\ref{fig:plot_v2_1} the experimental excitation function of $v_2$ is compared to the hydrodynamic model for different equations of
state and to the UrQMD calculation. The impact parameter in the simulations is $b=7$~fm.

For the UrQMD cascade simulation, one observes the correct trend of the elliptic
flow as a function of energy, namely an increase from negative to positive
values, however, the absolute values of the flow parameter $v_2$ are
significantly lower than the experimental data. This is consistent with the
findings of  \cite{Danielewicz:2002pu,Hillmann:2018nmd} where it was demonstrated that the additional
assumption of a hard equation of state (or a soft EoS with momentum dependence) is needed in
order to describe the elliptic flow with the microscopic UrQMD model. A
cascade of on-shell nucleons and secondaries scattering according to
unmodified cross sections apparently does not create the observed flow
characteristic \cite{Danielewicz:2002pu}. 

The results for the hydrodynamic simulation with the hadron gas equation of
state do not show a sign change of the elliptic flow parameter. In fact,
$v_2$ is positive at all energies and it is significantly higher than the
experimental data. The lack of the 'squeeze-out' effect at low
energies is plausible, since the hadron gas equation of state is very soft
at low values of energy and baryon density (see Fig.~\ref{fig:plot_eos}). Of all
hydrodynamic scenarios, the excitation function of the elliptic flow rendered by the chiral model 
equation of state exhibits the best agreement with the experimental curve.
The bag model equation of state, on the other hand, 'exaggerates' the actual
trend of the elliptic flow. The sign change is reproduced, but the absolute
negative and positive values are significantly larger. The reason for this behavior - as compared
to the chiral model - can be explained in terms of the equations of state
as shown in Fig.~\ref{fig:plot_eos}: The strong squeeze-out effect can be
attributed to the relative stiffness of the bag model equation of state in the
purely hadronic phase. The extreme softness of the bag model equation of
state in the mixed phase, on the other hand, is related to a delayed
expansion, which clearly favors positive elliptic flow due to the
absence of spectator matter during the expansion phase.

Again, we contrast the 'event-by-event' simulations with the calculations employing 'averaged' initial conditions, shown in Fig.~\ref{fig:plot_v2_1} as open symbols. As in the case of the
directed flow, the (positive or negative) elliptic flow effects in the 'averaged' scenario are stronger
than for the more realistic scenario with fluctuating initial state configurations. We conclude that fluctuating initial conditions and a realistic EoS with a transition to a QGP can best describe the flow data in this energy regime.

\section[]{Summary and conclusion}
We have revisited the one-fluid  description of nuclear collisions for the low energy regime. To this aim, we used 
a SHASTA implementation of ideal relativistic one-fluid hydrodynamics in order to
study central Au+Au reactions in the beam energy range between 1 and 10 $A$GeV. By applying a 
relativistically invariant particlization scheme with conservation of
average energy, net-baryon number and charge, we extracted  final state samples of individual hadrons. 
The resulting $m_t$ spectra of protons, $\pi^+$ and $K^+$ at midrapidity depend
to some extent on the assumed equation of state. Very good agreement
with experimental data is observed for the hadron gas equation of state and
$E_{\rm lab}$ between 6 and 10 $A$GeV. However, the spectra from the hydrodynamic
model scenarios are remarkably similar to the microscopic UrQMD model
results at all energies and for all particle species.
The sensitivity of the hydrodynamic model to the value of the local energy density $\epsilon_{\rm fr}$ which  defines the particlization hypersurface, is found to be very weak. 

First and second flow harmonics of protons were analyzed and found to be much more sensitive to the
Equation of State than inclusive particle spectra. The comparison to the experimental data indicates that the $v_2$ data can be best described with a chiral euqation of state that includes a phase transition to the QGP.
In addition, we explored the effect of single-event density fluctuations of the initial state on the
flow observables $v_1$ and $v_2$. We found that averaged initial stages do not allow for a realistic description of the flow data. 

In view of the above results we conclude that the relativistic one-fluid hydrodynamic
approach with particlization constitutes a useful baseline to predict and explore effects of the QCD phase transition in the FAIR/NICA energy range. Especially the improved EoS's, the fluctuating initial conditions and the state-of-the-art Monte Carlo freeze-out/transition procedure improve the results and reliability of the current one-fluid models over the their ancestors. 

\begin{acknowledgements}
This work was supported by the Helmholtz International Center for FAIR within the framework of the 
LOEWE program launched by the State of Hesse.
The computational resources were provided by the Center for Scientific Computing (CSC) of the 
Goethe University Frankfurt.
This work has been supported by COST Action THOR (CA15213). 
\end{acknowledgements}

\bibliography{myreferences}%

\end{document}